\documentclass[12pt,]{article}
\usepackage{amssymb}
\usepackage{graphicx}
\usepackage[cp1251]{inputenc}
\usepackage[russian]{babel}
\usepackage{rotating}
\begin{document}

 \centerline{\small\it
 Publications of the Pulkovo Observatory (ISSN 0367--7966), Issue 228,
 DOI:10.31725/0367-7966-2023-228-3}
 \bigskip
 \bigskip

 \centerline{\bf REVIEW OF CURRENT ESTIMATES OF THE GALAXY MASS}
 \bigskip
 \bigskip
 \centerline{\bf
  Vadim V. Bobylev\footnote [1]{vbobylev@gaoran.ru},
  Anisa T. Bajkova }

 \bigskip
 \centerline{\small\it
 Central Astronomical Observatory of the Russian Academy of Sciences, Pulkovo}

 \bigskip
 \bigskip

{\bf Abstract}---An overview of the methods used to estimate the mass of the Galaxy and the results obtained by various authors recently according to modern data is given. In particular, the estimates obtained based on the analysis of the galactic rotation curve, on the kinematics of the Galactic dwarf satellites and globular clusters, on the streams of such dwarf galaxies, on escape speed, as well as on halo stars are considered. Estimates of the Galaxy mass in the form $M (<r)$, $M_{\rm 200}$ and $M_{\rm vir}$ are considered.
According to 20 individual estimates, the average value was found
 $\overline M_{\rm 200}=0.88\times 10^{12}~M_\odot$ with a dispersion of
 $0.24\times 10^{12}~M_\odot$ and a weighted average error of $0.06\times 10^{12}~M_\odot$.
According to 25 individual estimates, $\overline M_{\rm vir}=1.02\times10^{12}~M_\odot$ was obtained with a dispersion of $0.41\times 10^{12}~M_\odot$ and a weighted average error of $0.09\times10^{12}~M_\odot$.

 \bigskip
\section*{Introduction}
The mass value is a key parameter for constructing a dynamic model of the Galaxy. Various methods are used to estimate this value. All of them are based on the analysis of the kinematics of stars and globular clusters belonging to the Galaxy, its surrounding dwarf satellite galaxies with their plumes, as well as neighboring galaxies (meaning the Andromeda nebula with its surroundings).

Of great importance in this task are the accuracy of determining the distances to the analyzed objects and their speeds. After all, the main objects for analysis are a variety of stars, for example, Cepheids, red giants, branch giants, RR~Lyr type variables, stars in globular clusters, stars in the plumes of globular clusters and satellite galaxies of the Milky Way.

Currently, the sources of the most accurate mass kinematic data are catalogs with measured trigonometric parallaxes and proper motions of stars. These are catalogs obtained as a result of space observations, such as Hipparcos (1997) and Gaia (Gaia Collab. 2016), measurements of the velocities of globular clusters and satellite galaxies of the Milky Way from the Hubble Space Telescope (for example, Libralato et al. 2018).

In Gaia\,EDR3 (Gaia Early Data Release 3, Gaia Collab. 2021), the trigonometric parallaxes of about 500 million stars were measured with errors of less than 0.2 milliarcseconds (mas). For stars with stellar magnitudes $G<15^m$, random errors in measuring proper motions lie in the range of 0.02--0.04 milliarcseconds per year (mas/yr). In general, the proper motions of about half of the catalog stars are measured with a relative error of less than 10\%. In Gaia\,EDR3, the radial velocity values were copied from the previous version of the catalog, from Gaia\,DR2 (Gaia Collab. 2018). A recently published version of the Gaia\,DR3 (Gaia Collab. 2022), where the radial velocities of stars are significantly improved compared to the velocities from Gaia\,DR2, and the values of parallaxes and proper motions of stars are simply copied from Gaia\,EDR3.

Distances to stars located from the center of the Galaxy further than $\sim$20~kpc are estimated photometrically. Only recently have all-celestial photometric surveys of the sky in the near infrared range appeared, which are extremely necessary for reliable accounting of interstellar absorption, such as WISE (Wide-field Infrared Survey Explorer, Wright et al. 2010) or GLIMPSE (Galactic Legacy Infrared Mid-plane Survey Extraordinaire, Benjamin et al. 2003), obtained in the result of space observations.

For these purposes, mass spectroscopic surveys of the sky are important, such as SDSS (Sloan Digital Sky Survey, York 2000), APOGEE (Apache Point Observatory Galactic Evolution Experiment, Eisenstein et al. 2011, Majewski et al. 2017), LAMOST (Large sky Area Multi-Object fiber Spectroscopic Telescope, Deng et al. 2012), RAVE (RAdial Velocity Experiment, Steinmetz et al., 2006), or GALAH (GALactic Archaeology with HERMES spectroscopic survey, Buder et al. 2021), containing information on the spectra and radial velocities of hundreds of thousands of stars.

The purpose of this work is to review the methods that are used to estimate the mass of the Galaxy. As well as an overview of the results obtained recently according to the most reliable data.

\section{Coordinate systems}
Directly from the observations we have the radial velocity $V_r$ and the components of the proper motion of the star $\mu_\alpha\cos\delta$ and $\mu_\delta,$ using which we can obtain two projections of the tangential velocity $V_l=4.74d\mu_l\cos b$ and $V_b=4.74d\mu_b,$ directed along the galactic longitude $l$ and latitude $b$ respectively, $d$~--- the heliocentric distance of the star in the kpc.

Through the components $V_r, V_l, V_b$, the velocities $U,V,W$ are calculated, directed along the rectangular galactic axes of the heliocentric coordinate system $x,y,z$. The velocity $U$ is directed from the Sun to the center of the Galaxy, $V$ in the direction of rotation of the Galaxy and $W$ to the north galactic pole. Two velocities: $V_R$, directed radially from the galactic center and the orthogonal velocity $V_{circ}$, directed along the rotation of the Galaxy, can be found based on the following relations:
\begin{equation}
 \begin{array}{lll}
  V_{circ}= U\sin \theta+(V_0+V)\cos \theta, \\
       V_R=-U\cos \theta+(V_0+V)\sin \theta,
 \label{VRVT}
 \end{array}
 \end{equation}
where the positional angle $\theta$ satisfies the relation $\tan\theta=y/(R_0-x)$, $V_0$~--- the linear rotation velocity of the Galaxy at the circumsolar distance $R_0$. The distance from the star to the axis of rotation of the Galaxy $R$ is calculated based on the ratio
 \begin{equation}
 R^2=d^2\cos^2 b-2R_0 d\cos b\cos l+R^2_0.
 \label{R}
 \end{equation}
In fact, the cylindrical coordinate system $R,\theta,z$ is given here.

We also use a rectangular galactocentric coordinate system $X,Y,Z$, in which the $X$ axis is directed from the center of the Galaxy to the Sun, the $Y$ axis is directed towards galactic rotation (this is the left coordinate system, but the right one is also often used), and the $Z$ axis is directed to the north pole of the Galaxy. Note that the coordinates $z$ and $Z$ are equivalent. In this coordinate system, the distance to the star will be denoted by $r$
 \begin{equation}
 r=\sqrt{X^2+Y^2+Z^2}.
 \label{rrr}
 \end{equation}

 \section{Methods for estimating the mass of a Galaxy}
We note the works of Karukes et al. (2020), Wang et al. (2020) devoted to this topic, which served as a model for us. These authors made an interesting review of the estimates of the mass of the Galaxy obtained by various authors, with a classification of methods for obtaining estimates. Here we continue this work with the addition of a number of new estimates.

\subsection{By the rotation curve}
The rotation curve is the dependence of the circular rotational velocities of stars $V_{\rm circ}$ around the axis of rotation of the Galaxy from the distance to its axis of rotation $R$. The galactic rotation curve is used to determine the parameters of a suitable model of the gravitational potential of the Galaxy $\Phi(R,z).$ Modern potential models are multicomponent, containing contributions from the main galactic subsystems, such as, for example, the central bulge, disk and halo. In this case $\Phi(R,z)=\Phi_b(r(R,z))+\Phi_d(r(R,z))+\Phi_h(r(R,z))$, where
$\Phi_b(r(R,z))$~--- bulge's contribution,
$\Phi_d(r(R,z))$~--- disk contribution and
$\Phi_h(r(R,z))$~--- contribution of the dark matter halo.
It is possible to specify the most common specific expressions in the potential model currently used to describe these subsystems.

1).~The Plummer sphere (Plummer 1911) models the galactic bulge:
\begin{equation}
  \Phi_b(r)=-\frac{G M_b}{\sqrt{r^2+b_b^2}},
  \label{bulge}
 \end{equation}

2).~A compressed spheroid is modeled as a disk in the form proposed in the paper
 Miyamoto, Nagai~(1975):
 \begin{equation}
 \Phi_d(R,z)=-\frac{G M_d}{\sqrt{ R^2+\biggl[ a_d+\sqrt{z^2+b_d^2} ~\biggr]^2  }},
 \label{disk}
\end{equation}

3).~The dark matter halo is modeled according to the work of Navarro et al. (1997), the approach is called the Navarro, Frank, White model:
\begin{equation}
  \Phi_h(r)=-\frac{G M_h}{r} \ln {\biggl(1+\frac{r}{a_h}\biggr)},
 \label{halo-III}
 \end{equation}
where $M_b, M_d, M_h$~--- component masses,
    $b_b, a_d, b_d, a_h$~--- scale parameters of the component.
In the works of Bobylev, Bajkova (2016; 2017~a,b) and Bobylev et al. (2017), the approach based on the relations (\ref{disk})--(\ref{halo-III}) corresponds to the model~III.

Knowing the parameters of the potential $\Phi(R,z),$ it is possible to estimate the mass of the Galaxy contained in a sphere of radius $r$
\begin{equation}
M(<r)=\frac{r^2}{G}\frac{d\Phi(r)}{dr},
\label{m}
\end{equation}
where $G$~--- gravitational constant,
mass $M(<r)$ in the future we will denote by $M_{\rm r}$.

Note that in the simplest case of a spherically symmetric distribution of stellar density, the mass of matter in a sphere with radius $r$ can be estimated by the formula:
\begin{equation}
M(<r)=\frac{r V^2}{G},
\label{m-555}
\end{equation}
which can be obtained by equating the centrifugal force with the force of attraction.

The axisymmetric model of the gravitational potential of the Galaxy in work of Zhou et al. (2022) included the following four components: bulge, thin and thick disks, as well as a halo of dark matter. Kinematic data on approximately 54,000 stars belonging to the branch of red giants were used to construct the rotation curve of the Galaxy. Moreover, these data were taken from the Gaia\,EDR3 catalog. Spectral information was taken from APOGEE and LAMOST surveys, and near-infrared photometry from the 2MASS catalog was also used (Skrutskie et al. 2006). In Zhou et al. (2022), an estimate of the mass of the Galaxy was obtained
$M_{\rm vir}=(0.805\pm0.115)\times10^{12}$~M$_\odot$ with the found value of the virial radius $r_{\rm vir}=192.37\pm9.24$~kpc.

The virial mass $M_{\rm vir}$ is defined as the mass enclosed within the virial radius of a gravitationally bound system $r_{\rm vir}$, where $r_{\rm vir}$ is the radius within which the system obeys the virial theorem, according to which the system is in equilibrium with the external environment if the doubled value of the averaged kinetic $T$ system is equal to the potential
$U:~2\langle T\rangle=-\langle U\rangle$.

Sometimes the following method of estimating $M_{\rm vir}$ is used
(Xue et al., 2008; Huang et al., 2016; Bird et al., 2022):
\begin{equation}
 M_{\rm vir}=\frac{4\pi}{3}\rho_{\rm crit} \Omega_m \delta_{\rm th} r^3_{\rm vir},
\label{m-VIR}
\end{equation}
where $\rho_{\rm crit}=3H^2_0/8\pi G$ is the critical density of the Universe (with which the density of the Galaxy should be compared), $\Omega_m$ is the contribution of visible matter to the critical density, $\delta_{\rm th}$ is critical super-density during virialization. Xue et al. (2008) accepted $\Omega_m=0.3$, $\delta_{\rm th}=340,$$H_0=65$~km/s/Mpc. Thus, the estimate of $M_{\rm vir}$ (which ever way it is calculated) strongly depends on the accepted model.

In Ablimit et al. (2020), the rotation curve of the Galaxy is constructed from a sample of 3\,500 classical Cepheids. Photometric data in the infrared range were collected for all these Cepheids and distance estimates were obtained using the period-Wesenheit ratio according to the calibrations of Wang et al. (2018). For kinematic analysis, however, 1078 Cepheids with measured radial velocities were selected. As a result, the parameters of the gravitational potential of the Galaxy were refined and a new virial estimate of the mass of the Galaxy was obtained, $M_{\rm vir}=(0.822\pm0.052)\times 10^{12}~M_\odot$, where the value of the virial radius of the Galaxy was found to be $r_{\rm vir}=191.84\pm4.12$~kpc.

Eilers et al. (2019) constructed a rotation curve of the Galaxy using $\approx$25\,000 red giants located in the distance range $R:5-25$~kpc. Kinematic data were taken from the Gaia\,DR2 catalog, spectral information from the APOGEE survey, and near-infrared photometry from the 2MASS catalog. Eilers et al. (2019) obtained an estimate of the virial mass of the Galaxy, which was
$M_{\rm vir}=(0.725\pm0.025)\times 10^{12}~M_\odot$. These authors do not report on the virial radius of the Galaxy, while it is obvious that its value in their model differs little from the estimates of Zhou et al. (2022) and Ablimit et al. (2020).

Bhattacharjee et al. (2014) constructed the rotation curve of the Galaxy from data extending up to $R\sim200$~kpc. For this purpose, the kinematic characteristics of a wide variety of objects were used. In particular, hydrogen clouds, scattered star clusters, OB stars, Cepheids, carbon stars, various giants, globular clusters and dwarf satellite galaxies of the Milky Way participated. This is not a complete list yet. Bhattacharjee et al. (2014) received an assessment
 $M_{\rm 200}=(0.68\pm0.41)\times 10^{12}~M_\odot$.

Bajkova, Bobylev (2016; 2017) and Bobylev et al. (2017) added masers with trigonometric parallaxes measured by the VLBI method to the process of constructing the rotation curve of the Galaxy, and data from Bhattacharjee et al. (2014) were used for a more distant interval. Six models of galactic potential were tested. Note that from the rotation curve of the Galaxy, plotted from the velocities of objects in the entire range $R:0-200$~kpc, we obtain a more reliable estimate of the mass of $M_{\rm 200}$. Compared to what is obtained as a result of extrapolation. Based on the most suitable model (Model III), we obtained an estimate
$M_{\rm 200}=(0.75\pm0.19)\times 10^{12}~M_\odot$.

For very distant objects (further $\approx$50~kpc from the center of the Galaxy), such as globular clusters and dwarf galaxies, it is difficult to find the circular rotational velocities of $V_{circ}$ directly by the formula~(\ref{VRVT}). To do this, indirect estimates of the velocity $V_{circ}$ are used, based on the application of the Jeans equation, through the dispersions of the radial velocities of stars. One of the variants of the approach (Binney, Tremaine 1987) looks like this:
\begin{equation}
  V^2_{circ} (R) = -\frac{r}{\rho} \frac{d(\rho \sigma^2_r)}{dt} -2\beta\sigma^2_r,
  \qquad\quad \beta=1-\frac{\sigma^2_t}{\sigma^2_r},
  \label{Jeans-1}
 \end{equation}
where $\sigma_r(r)$ and $\sigma_t(r)$ are the radial and tangential velocity dispersions, respectively, $\rho(r)$ is the stellar density, and $\beta$ is the anisotropy parameter. This approach, for example, is used in the work of Xue et al. (2008) to calculate the circular velocities of the blue giants of the horizontal branch, where they found $M_{\rm 60}=(0.40)\pm0.07)\times 10^{12}~M_\odot$.

In Huang et al. (2016), the rotation curve of the Galaxy is constructed from stars that quite densely fill the distance interval up to $\sim100$~kpc. To do this, we used data on HI hydrogen clouds in the inner part of the Galaxy, about 16\,000 red-clump stars, as well as about 5\,700 halo giants of spectral class K. As a result, these authors received an assessment
 $M_{\rm vir}=(0.90^{+0.07}_{-0.08})\times 10^{12}~M_\odot$
with the found value $r_{\rm vir}=255.7^{+7.7}_{-7.7}$~kpc.
To calculate the circular velocities of distant K-giants, these authors also had to use an approach based on the Jeans equation~(\ref{Jeans-1}).

It should be noted that McMillan (2017) constructed a model of the potential of the Galaxy only from data on masers with trigonometric parallaxes measured by the VLBI method and found
 $M_{\rm 100}=(0.82\pm0.11)\times 10^{12}~M_\odot$.

Sofue (2012) constructed a rotation curve of the Galaxy, covering a huge range of distances -- up to $R\sim1$~Mpc. Data on the velocities of galactic globular clusters and dwarf galaxies surrounding the Milky Way and the Andromeda nebula were involved. The estimated mass of the Galaxy was $M_{\rm 365}=(0.703\pm0.101)\times 10^{12}~M_\odot$, where the value of the midpoint of the distance between the Milky Way and the Andromeda nebula was chosen as the radius of the enclosing sphere $r=365$~kpc. To estimate the mass of the Galaxy, an ``honest" integration of the density of matter was carried out in accordance with the selected model of the potential of the Galaxy.

\subsection{By globular clusters of the Milky Way}
Globular clusters are very important for analysis. For them, it is possible to calculate the average values of distances and speeds with relatively high accuracy. Globular clusters belong to the spherical component of the Galaxy. They are distributed in a huge area of the Galaxy, then there are only dwarf galaxies-satellites of the Milky Way. Therefore, globular clusters are often used to plot the rotation curve of the Galaxy both in combination with other data and separately.

Sohn et al. (2018) analyzed 20 galactic globular clusters. The values of their average proper motions were calculated from observations of the Hubble Space Telescope. These authors found
 $M_{\rm 39.5}=(0.61^{+0.18}_{-0.12})\ times 10^{12}~M_\odot$ and
 $M_{\rm vir}=(2.05^{+0.97}_{-0.79})\times 10^{12}~M_\odot$.

Posti and Helmi (2019) analyzed the proper motions of 75 globular clusters, which were calculated from data from the Gaia\,DR2 catalog. Then 20 more distant globular clusters were included in the sample, the values of the average proper motions of which were obtained using the Hubble Space Telescope. These authors found
 $M_{\rm 20}=(0.191^{+0.018}_{-0.017})\times 10^{12}~M_\odot$.

Watkins et al. (2019) used 34 globular clusters for analysis. The calculation of the kinematic characteristics of these clusters was made according to the data of the Gaia\,DR2 catalog and observations from the Hubble Space Telescope. These authors found $M_{\rm 39.5}=(0.42^{+0.07}_{-0.06})\times 10^{12}~M_\odot$.

In Vasiliev (2019), according to Gaia\,DR2 data, the average proper motions of 150 globular clusters were calculated. Currently, this is the most massive catalog of globular clusters with known estimates of distances, proper motions and radial velocities. From the analysis of the rotation curve constructed on the basis of a two-component (disk and halo) potential model, he obtained an estimate
 $M_{\rm 100}=(0.85^{+0.33}_{-0.20})\times 10^{12}~M_\odot$.

In Wang, et al. (2022), the rotation curve of the Galaxy is constructed using the average proper motions of 150 globular clusters, which they calculated according to the data of the Gaia\,EDR3 catalog. These authors concluded that the total mass of the Galaxy is from $M_{\rm 200}=(0.536^{+0.081}_{-0.068})\times 10^{12}~M_\odot$
up to $M_{\rm 200}=(0.784^{+0.308}_{-0.197})\times 10^{12}~M_\odot$ depending on the halo model. Moreover, these estimates are made taking into account the contribution of the Large Magellanic Cloud (BMO). That is, taking into account the contribution of the BMO at a distance interval of $<100$~kpc slightly reduces the analyzed circular velocities of objects. Therefore, these authors talk about the underestimated values of the estimates of $M_{\rm 200}$ in comparison with the known ones.

In the work of Sun et al. (2023), the rotation curve of the Galaxy is constructed from 159 globular clusters. Their average proper motions, calculated according to the Gaia\,EDR3 catalog data, were taken from the work of Vasiliev, Baumgardt (2021). Sun et al. (2023) found $M_{\rm 200}=(1.11^{+0.25}_{-0.18})\times 10^{12}~M_\odot$.

Tidal tails or stellar plumes are formed as a result of the destruction of scattered star clusters, globular clusters and dwarf satellite galaxies of the Milky Way. When such plumes have a large extent, they actually plot the history of the movement of a cluster or dwarf galaxy over a long period of time in the field of attraction of the Galaxy. The analysis of such plumes provides another dynamic method for estimating the mass of the Galaxy.

Note the work of K\"upper et al. (2015), in which the plume from the globular cluster Palomar~5 was studied. These authors obtained an estimate inside a sphere with a radius of 19 kpc, which is equal to the radius of the apogalactic orbit of the Palomar cluster~5:
$M_{\rm 19}=(0.21\pm0.04)\times 10^{12}~M_\odot$, and even appreciated
 $M_{\rm 200}=(1.69\pm0.42)\times 10^{12}~M_\odot$.

 \subsection{On the kinematics of satellite galaxies}
Almost all authors who analyze the galactic orbits of dwarf satellite galaxies of the Milky Way note that the estimates of the mass of the Galaxy obtained from these objects strongly depend on the Leo~I galaxy (for example, Boylan-Kolchin et al. 2013). This is the most distant ($d=261$~kpc) of the dwarf satellite galaxies of the Milky Way, even using measurements of its own motion from a space telescope (Sohn et al. 2013) does not yet allow us to say unequivocally (Bajkova, Bobylev 2017) whether its orbit is hyperbolic or elliptical, i.e., it is connected gravitationally with the Galaxy or not.

Boylan-Kolchin et al. (2013) analyzed the motion of the dwarf galaxy Leo~I using measurements of its own motions obtained using the Hubble Space Telescope. As a result, these authors received an estimate of the mass of the Galaxy
 $M_{\rm vir}=1.6\times 10^{12}~M_\odot$.

Eadie et al. (2015) investigated a combined sample of globular clusters and dwarf satellite galaxies of the Milky Way. These authors obtained an estimate of the mass of the Galaxy $M_{\rm 260}=(1.37)\pm0.12)\times 10^{12}~M_\odot$.

Fritz et al. (2020) analyzed 45 dwarf satellite galaxies of the Milky Way. The proper motions and radial velocities of stars for calculating the average values were taken from the Gaia\,DR2 catalog. These authors obtained an estimate of the value of the virial mass of the Galaxy
 $M_{\rm 308}=(1.51^{+0.45}_{-0.40})\times 10^{12}~M_\odot$.

 \subsection{By the plumes of satellite galaxies}
Quite a lot of plumes are known in the Galaxy, formed as a result of the destruction of globular clusters and dwarf galaxies. A detailed overview of such structures can be found in Grillmair, Carlin (2016). Well-known, for example, such names as Sagittarius stream, GD-1 or Orphan Stream.

Gibbons et al. (2014) studied the plumes formed by a dwarf satellite galaxy of the Milky Way in Sagittarius (Sagittarius stream). These authors received an assessment
$M_{\rm 100}=(0.41\pm0.04)\times 10^{12}~M_\odot$, and extrapolation to a more distant distance gave $M_{\rm 200}=(0.56\pm0.12)\times 10^{12}~M_\odot$.

In Malhan, Ibata (2019), data on the GD-1 plume were used to refine the model of the gravitational potential of the Galaxy. As a result, these authors found
$M_{\rm 20}=(0.25\pm0.2)\times 10^{12}~M_\odot$.

 \subsection{According to the speeds of departure from the Galaxy}
The values of the escape velocities from the Galaxy, or escape velocities from its gravitational field, depend on the distance $R$. The analysis of such velocities makes it possible to refine the model of the gravitational potential of the Galaxy. In this case, the expression~(\ref{m}) to estimate the mass of the Galaxy can be transformed to the following form (Williams et al. 2017):
\begin{equation}
M(<r)=\frac{-r^2 V^2_{\rm esc}} {G} \frac{dV_{\rm esc}}{dr},
\label{m-Vesc}
\end{equation}
where $V_{\rm esc}$ is escape velocity.

Williams et al. (2017) the observed distribution of the radial velocities of stars ($v_{\rm ||}$) on the plane $r-v_{\rm||}$ and the limitations of the model found by them were investigated. These authors used three stellar samples of halo indicators: MSTO (main-sequence turn-off stars), K-giants and BHBs (blue stars of the horizontal branch) according to data from the SDSS catalog (Ahn et al. 2012). MSTO stars are numerous, but mostly observed at distances up to $\approx3$~kpc from the Sun. Statistical data on these stars allow us to find limits on the value of the escape velocity in the region of the Sun. The number of K-giants and BHBs is smaller, but they are bright and were observed at distances of $\approx.50$~kpc from the Sun, which increased the spatial extent of the sample used to the range of $R\approx.40$~kpc. As a result, Williams et al. (2017) found the value $M_{\rm 50}=(0.30^{+0.07}_{-0.05})\times 10^{12}~M_\odot.$

\subsection{By the stars of the halo}
Deason et al. (2012) analyzed blue horizontal branch stars with measured radial velocities distributed in the region of $R>80$~kpc. These authors found $M_{\rm 150}: [0.5-1.0]\times 10^{12}~M_\odot.$

Williams, Evans (2015) conducted an interesting study of the velocities of halo stars. In Fig.~6 of the work of Williams, Evans (2015), ellipsoids of the velocities of halo stars are shown. Ellipses are colored depending on the parameter value $\beta=1-(\sigma^2_\phi+\sigma^2_\theta)/2\sigma^2_r,$ which is a measure of the local velocity anisotropy (here the angle $\phi$ is counted from the galactic plane). If $\beta>0~(<0),$ then the model has a radial (tangential) displacement, and $\beta=0$ implies an isotropic model. As a result, Williams, Evans (2015) concluded that $M_{\rm 50}\approx0.45\times 10^{12}~M_\odot.$

In Ablimit, Zhao (2017), 860 variables of the RR~Lyr type belonging to the stellar halo were used to construct the galactic rotation curve. High-precision distances and radial velocities were determined for these stars. They are distributed in the range $r:0-60$~kpc. The circular velocities were estimated based on the Jeans equation. And based on the ratio~(\ref{m-555}), the mass of the Galaxy was found
$M_{\rm 50}=(0.375\pm0.133)\times 10^{12}~M_\odot.$

Bird et al. (2022) on the K-giant halos found
 $M_{\rm 200}=(0.55^{+0.15}_{-0.11})\times 10^{12}~M_\odot,$
a by the blue horizontal branch stars
 $M_{\rm 200}=(1.00^{+0.67}_{-0.33})\times 10^{12}~M_\odot.$

Among the halo stars, the so-called hypervelocity stars were discovered relatively recently, flying from the center of the Galaxy at speeds of $\approx700$~km/s (Brown et al. 2010). That is, these are fleeing stars, so they make it possible to estimate the mass of the Galaxy. In the work of Gnedin et al. (2010), using a sample of hypervelocity halo stars, an estimate of the virial mass of the Galaxy was obtained
 $M_{\rm vir}=(1.6\pm0.3)\times10^{12}~M_\odot$ for the found value
 $r_{\rm vir}=300$~kpc. According to data on hypervelocity stars Fragione, Loeb (2017) found that the mass of the Galaxy is enclosed in the range $[1.2-1.9]\times10^{12}~M_\odot.$

\subsection{By equilibrium with the Andromeda nebula}
There are methods for estimating the mass of a Galaxy without plotting its rotation curve. For example, Karachentsev et al. (2009) received an estimate of the total mass of the Local Group $M_{\rm LG}=(1.9\pm0.2)\times10^{12}~M_\odot.$ The ratio of the masses of the Galaxy and M31 was 4:5, i.e., M31 with its surroundings is slightly more massive than the Milky Way system. For this assessment, the effect of deceleration of the local Hubble flow and data on the distances and radial velocities of galaxies in the vicinity of the Local Group were used. The total mass of the Galaxy obtained by this independent method is  $M_{350}=(0.84\pm0.09)\times10^{12} M_\odot$.

There is currently no agreement on the estimate of the total mass of the Local Group of Galaxies. For example, in van der Marel et al. (2012), found
$M_{\rm LG}=M {(\rm MW)}_{\rm vir} + M {(\rm M31)}_{\rm vir}=(3.17\pm0.57)\times10^{12} M_\odot$ involving cosmic measurements of the proper motions of stars in the galaxy M31. Here, about half of the mass is accounted for by the mass of the Milky Way.

Gonz\'alez et al. (2014) from the analysis of measurements of the proper motion of the galaxy M31 found
$M {(\rm MW)}_{\rm 200} + M {(\rm M31)}_{\rm 200}=(2.40^{+1.95}_{-1.05})\times10^{12} M_\odot$, which is in better agreement with the individual estimates of the mass of the Milky Way described by us.

From the analysis of the rotation curve of the Galaxy Sofue (2009), as well as, for example, Karachentsev et al. (2009), Watkins et al. (2010), concluded that the mass of the Andromeda nebula slightly exceeds the mass of the Milky Way. But Pe\~narrubia et al. (2014), although they received a consistent estimate of the mass of the Local Group, $M{(\rm MW)}_{\rm 200} + M {(\rm M31)}_{\rm 200}=(2.3\pm0.7)\times10^{12}M_\odot$, claim that the mass of the Galaxy slightly exceeds the mass of the Andromeda nebula, since their mass ratio is $0.54^{+0.23}_{-0.17}$.

Finally, recently Carlesi et al. (2022) applied modeling and machine learning methods to study the velocities of the Milky Way and the Andromeda nebula within the framework of the standard $\Lambda$CDM cosmological model. As a result, they found
  $M_{\rm LG}=(3.31^{+0.79}_{-0.67})\times10^{12} M_\odot,$
  $M_{\rm MW}=(1.15^{+0.25}_{-0.22})\ times10^{12} M_\odot$ and
 $M_{\rm M31}=(2.01^{+0.65}_{-0.39})\ times10^{12} M_\odot.$
The mass ratio is $M_{\rm M31}/M_{\rm MW}=1.75^{+0.54}_{-0.28}$, where the mass of the Andromeda nebula is almost twice the mass of the Milky Way.

\begin{figure}[t]
    \centering
    \includegraphics[width=0.75\textwidth]{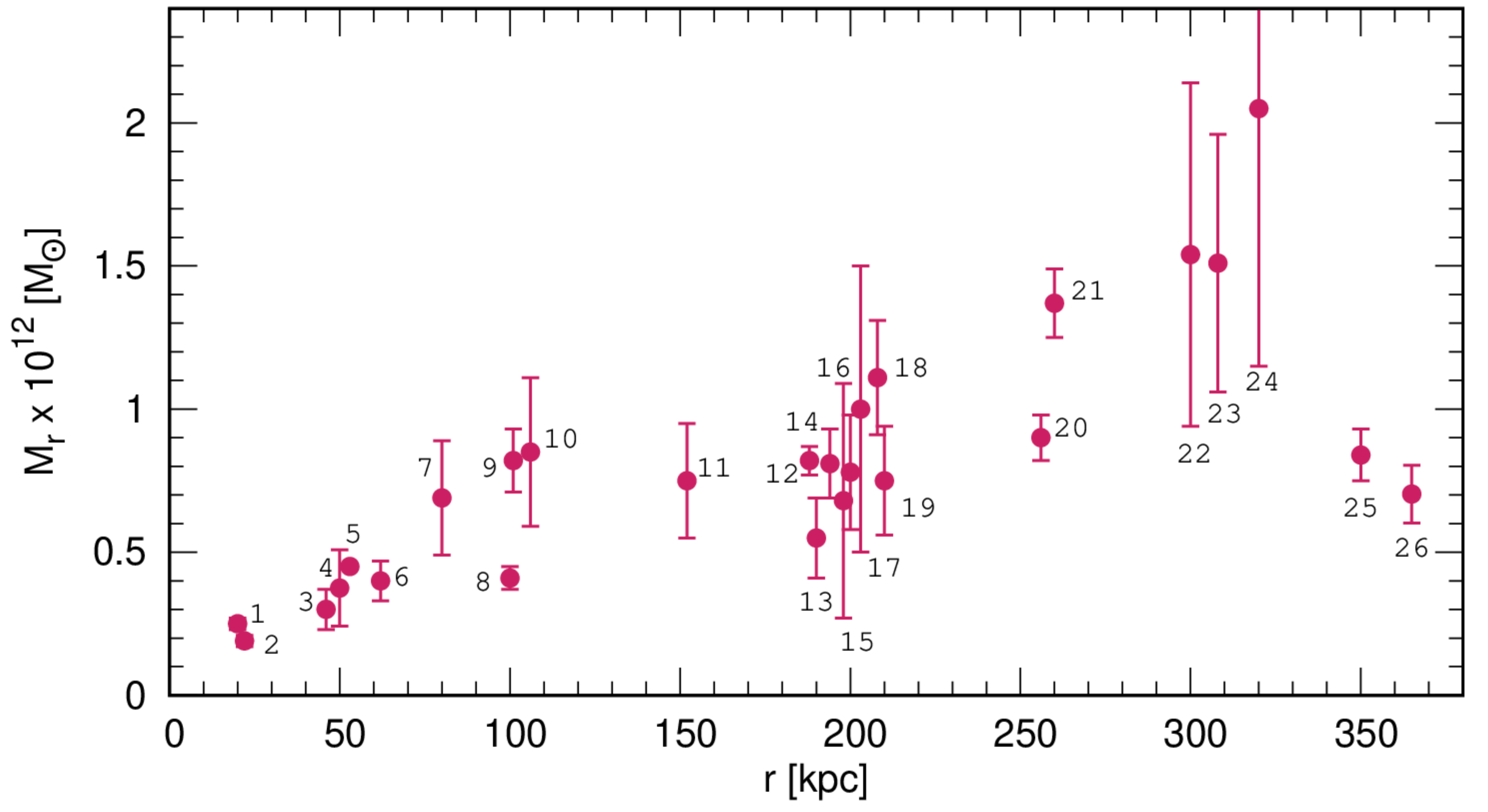}
    \caption{
Estimates of the mass of a Galaxy of the form $M_{\rm r}$,
1~--Malhan, Ibata (2019); 2~--Posti, Helmi (2019); 3~-- Williams et al. (2017);  4~-- Ablimit, Zhao (2017); 5~-- Williams, Evans (2015); 6~-- Xue et al. (2008);  7~-- Gnedin et al. (2010); 8~-- Gibbons et al. (2014);  9~-- McMillan (2017); 10~-- Vasiliev (2019); 11~-- Deason et al. (2012);  12~-- Ablimit et al. (2020); 13~-- Bird et al. (2022); 14~-- Zhou et al. (2022);  15~-- Bhattacharjee et al. (2014); 16~-- Wang, et al. (2022); 17~-- Bird et al. (2020); 18~-- Sun et al. (2023); 19~-- Bobylev, Bajkova (2016); 20~-- Huang et al. (2016);  21~-- Eadie et al. (2015); 22~-- Watkins et al. (2010); 23~-- Fritz et al. (2020);  24~-- Sohn et al. (2018); 25~-- Karachentsev et al. (2009); 26~-- Sofue (2012).
}
    \label{f-Mass-22}
\end{figure}
\begin{figure}[t]
    \centering
    \includegraphics[width=0.75\textwidth]{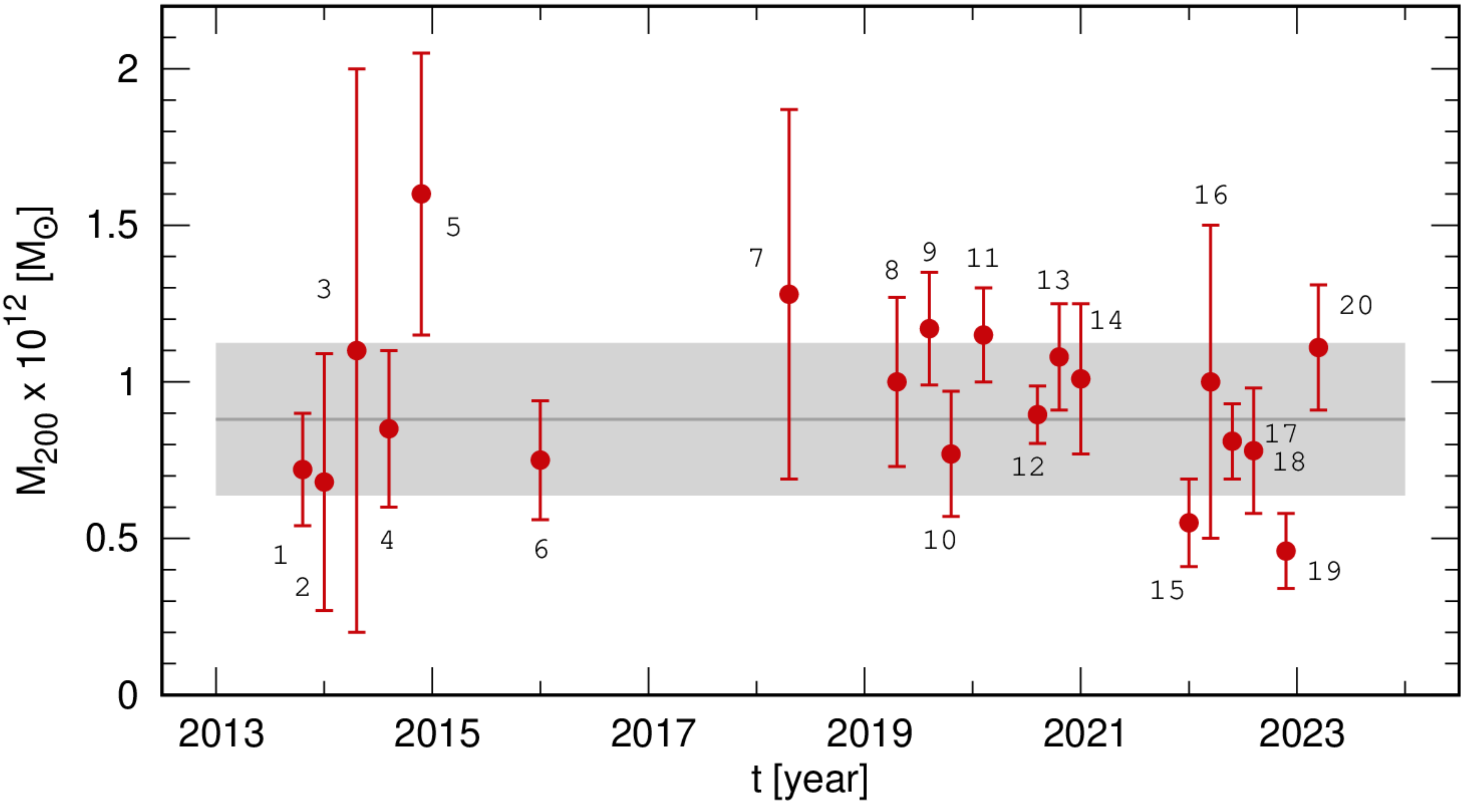}
    \caption{
Estimates of the mass of a Galaxy of the form $M_{\rm 200}$ depending on the year of publication,
1~-- Kafle et al. (2014); 
  2~-- Bhattacharjee et al. (2014); 
  3~-- Barber et al. (2014);    
  4~-- Cautun et al. (2014);    
  5~-- Piffl et al. (2014);     
  6~-- Bobylev, Bajkova (2016); 
  7~-- Monari et al.  (2018);   
  8~-- Deason at al.  (2019);   
  9~-- Callingham et al. (2019);
 10~-- Eadie, Juri\'c (2019);
 11~-- Li et al. (2020);     
 12~-- Karukes et al. (2020);
 13~-- Cautun et al. (2020); 
 14~-- Deason et al. (2021); 
 15~-- Bird et al. (2022);   
 16~-- Bird et al. (2022);   
 17~-- Zhou et al.   (2022); 
 18~-- Wang et al.   (2022); 
 19~-- Necib, Lin    (2022); 
 20~-- Sun, et al.   (2023). 
}
    \label{f-M-200}
\end{figure}
\begin{figure}[t]
    \centering
    \includegraphics[width=0.75\textwidth]{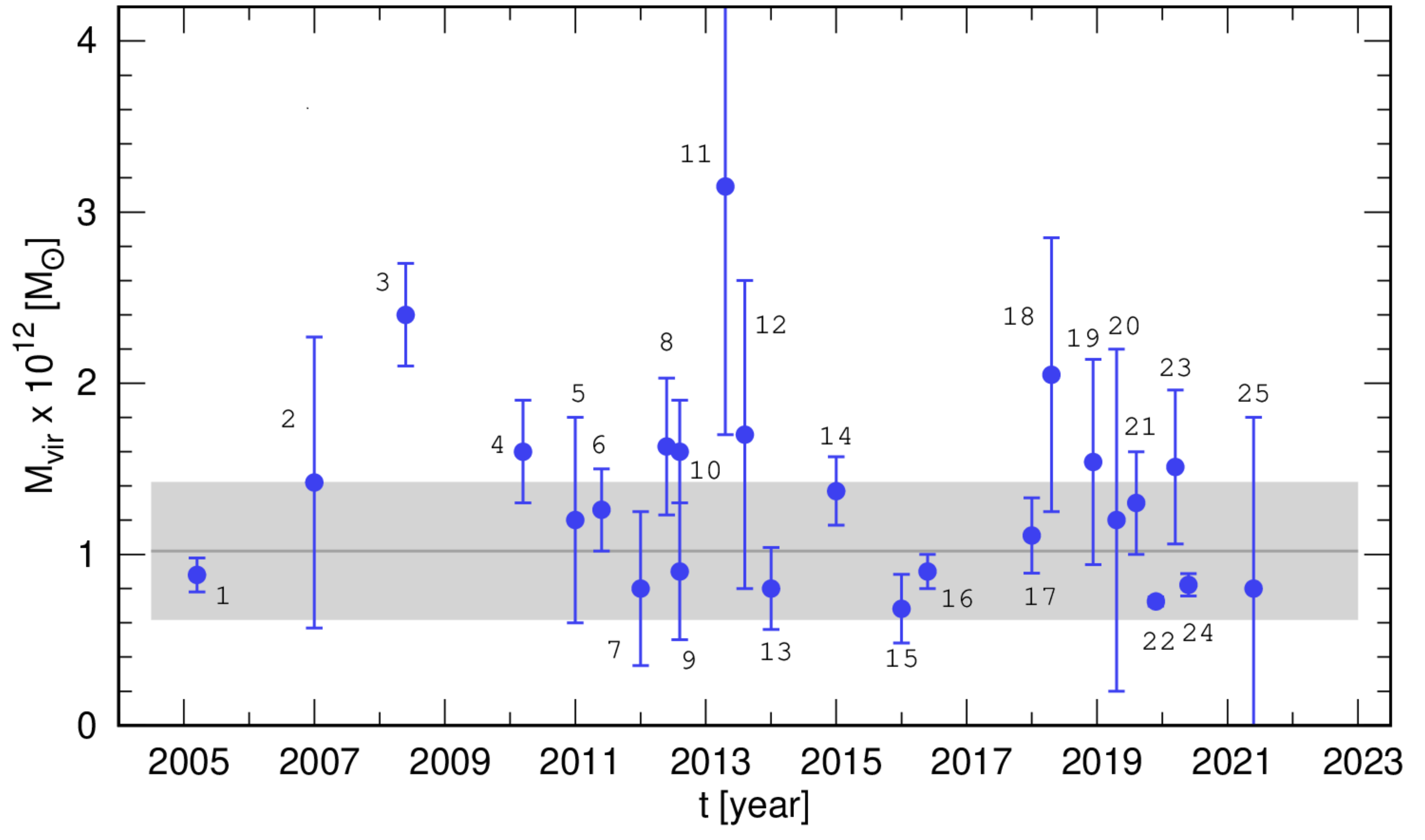}
    \caption{
Estimates of the mass of a Galaxy of the form $M_{\rm vir}$ depending on the year of publication,
1~-- Battaglia et al. (2005);
  2~-- Smith et al. (2007);
  3~-- Li, White (2008);
  4~-- Gnedin et al. (2010);
  5~-- Busha et al. (2011);
  6~-- McMillan (2011);
  7~-- Bovy et al. (2012);
  8~-- van der Marel et al. (2012);
  9~-- Kafle et al. (2012);
  10~-- Boylan-Kolchin et al. (2013);
  11~-- Sohn et al. (2013);
  12~-- Rashkov et al. (2013);
  13~-- Kafle et al. (2014);
  14~-- Eadie et al. (2015);
  15~-- Eadie, Harris (2016);
  16~-- Huang et al.(2016);
  17~-- Zhai et al. (2018);
  18~-- Sohn et al. (2018);
  19~-- Watkins et al. (2019);
  20~-- Vasiliev (2019);
  21~-- Posti, Helmi (2019);
  22~-- Eilers et al. (2019);
  23~-- Fritz et al. (2020);
  24~-- Ablimit et al. (2020);
  25~-- Li et al. (2021).
  }
    \label{f-M-vir}
\end{figure}

\section{Graphs with results}
The main results described in the previous section are reflected in Fig.~\ref{f-Mass-22}, where estimates of the mass of the Galaxy inside a certain radius~$r$ are given. In the range $r:0-200$~kpc, estimates naturally increase with increasing distance. At $r>200$~kpc, estimates of the mass of the Galaxy are mostly virial, only the most interesting ones are given. For a larger number of virial estimates, a separate figure will be given.

There is a sufficient number of estimates of the form $M_{\rm 200}$ for separate consideration. Obtained by integrating the rotation curve (see formula~(\ref{m})), they are more rigorous, compared to the estimates of $M_{\rm vir}$. Following the approach of Karukes et al. (2020), we constructed Fig.~\ref{f-M-200}, which displayed only the estimates of $M_{\rm 200}$. These estimates are taken on the condition that they were found using the rotation curve of the Galaxy and the authors accurately indicate $M_{\rm 200}$, without identifying it with the virial one.

It is known (for example, Bland-Hawthorn, Gerhard 2016) that the estimates $M_{\rm 200}$ and $M_{vir}$ are related by the ratio $M_{\rm 200}\approx0.85\times M_{vir}.$ It is interesting for us to repeat the calculation of averages for advanced statistics, compare the variances of estimates. To do this, the average $\overline M_{\rm 200}$ and the weighted average were calculated using weights inversely proportional to the errors of individual estimates $w=1/\sigma$, $(\overline M_{\rm 200})_w:$
 \begin{equation}
 \begin{array}{lll}
   ~~~\overline M_{\rm 200} = (0.94\pm0.06)\times10^{12}~M_\odot, \\
  (\overline M_{\rm 200})_w = (0.88\pm0.06)\times10^{12}~M_\odot,
 \label{mean-1}
 \end{array}
 \end{equation}
where the errors of the mean and weighted average are indicated, respectively, and the $\sigma$ variances of these two estimates are $0.27\times10^{12}~M_\odot$ and $0.24\times10^{12}~M_\odot$. In Fig.~\ref{f-M-200} the dark line shows the average
$(\overline M_{\rm 200})_w=0.88\times10^{12}~M_\odot$,
and the fill is given a confidence area corresponding to the error level of 68\%
(level $1\sigma=0.24\times10^{12}~M_\odot$).

Some authors give two estimates $M_{\rm r}$ and $M_{vir}$. Part of such estimates $M_{\rm r}$ is given in Fig.~\ref{f-Mass-22}. In Fig.~\ref{f-M-vir} 25 estimates of $M_{vir},$ obtained by various methods from various data are given. As can be seen from this figure, the errors of individual estimates and the spread of results are greater here compared to the data in Fig.~\ref{f-M-200}. The average values of ${\overline M_{\rm vir}}$ and $(\overline M_{\rm vir})_w$ calculated from these data are as follows:
\begin{equation}
 \begin{array}{lll}
   ~~~\overline M_{\rm vir} = (1.33\pm0.12)\times10^{12}~M_\odot, \\
  (\overline M_{\rm vir})_w = (1.02\pm0.09)\times10^{12}~M_\odot,
 \label{mean-2}
 \end{array}
 \end{equation}
where the errors of the mean and weighted average are indicated, respectively, and the variances of the estimates, $\sigma$, are $0.58\times10^{12}~M_\odot$ and $0.41\times10^{12}~M_\odot$, respectively. We see less agreement between the two estimates and their large errors compared to the result~(\ref{mean-1}).

\section{Discussion}
However, not everyone agrees that the mass of the Galaxy is $\leq1\times10^{12}~M_\odot.$
For example, Zaritsky, Courtois (2017) modeled analogues of the Milky Way, for which they selected measurements of gas mass in stars and disks from the literature, measurements of gas masses in halos, and also used a well-established value of the cosmological fraction of baryons. Thus, they estimated the lower bound of the mass of the Galaxy in a way independent of dynamics. As a result, these authors reject estimates of the low mass of the Galaxy ($\leq1\times10^{12}~M_\odot$), since such values imply a fraction of galactic baryonic matter significantly exceeding the global value. In their opinion, the convergence between dynamic mass estimates and estimates based on the baryon mass is an important milestone in understanding the evolution of galaxies.

In their review Bland-Hawthorn, Gerhard (2016) found the following average values of the mass of the Galaxy
 $\overline M_{\rm 200} = (1.1\pm0.3)\times10^{12}~M_\odot$ and
 $\overline M_{\rm vir} = (1.3\pm0.3)\times10^{12}~M_\odot$ inside a sphere with radius $r_{\rm vir} = 282\pm30$~kpc.
It can be seen that these results do not contradict our (\ref{mean-1}) and (\ref{mean-2}), and within the margin of error, the averages are in good agreement.

 \section*{Conclusion}
Estimates of the mass of the Galaxy obtained by various authors by the following most commonly used dynamic methods are considered:
a)~from the analysis of the galactic rotation curve, b)~on the kinematics of dwarf satellite galaxies of the Milky Way and globular clusters, c)~by the plumes of dwarf galaxies, d)~by escape velocity, e)~by the distant giants of the halo, as well as g)~on deceleration of the local Hubble flow.

Note that in the range $R:5-25$~kpc there is an excellent agreement of the Galactic rotation curves constructed by various authors according to modern kinematic data. There is good agreement in estimating the mass of a Galaxy of the form $M_{\rm r}$ up to distances 0--150~kpc, where it is currently possible to construct a rotation curve of the Galaxy from real objects.

It seems to us that such methods as estimating the mass of a Galaxy by the long (hereinafter $\sim100$~kpc) rotation curve of the Galaxy, analysis of the galactic orbits of distant globular clusters and dwarf satellite galaxies of the Milky Way or analysis of the escape velocities of stars are especially valuable. I.e., methods using the gravitational potential of the Galaxy based on analysis of the orbits of stars, globular clusters and dwarf galaxies. Methods that allow you to directly estimate the value of $M_{\rm 200}$.

Graphs with estimates of the mass of the Galaxy of the form $M_{\rm r}$ are constructed, estimates of the form $M_{\rm 200}$ and $M_{\rm vir}$ are considered separately. It is shown that the current estimates of the total mass of the Galaxy $M_{\rm vir}$ lie in the range $[0.5-2.5]\times 10^{12}~M_\odot,$
and estimates $M_{\rm 200}$~--- in the range $[0.4-1.6]\times 10^{12}~M_\odot.$

A weighted average value was found for 20 individual estimates
 $\overline M_{\rm 200}=0.88\times10^{12}~M_\odot$ with a variance of $0.24\times10^{12}~M_\odot$ and a weighted average error of $0.06\times10^{12}~M_\odot$.
According to 25 individual virial estimates , the same was obtained
 $\overline M_{\rm vir}=1.02\times10^{12}~M_\odot$ with variance
 $0.41\times10^{12}~M_\odot$ and a weighted average error of $0.09\times10^{12}~M_\odot$.

 \medskip
The authors are grateful to the reviewer for useful comments that contributed to the improvement of the manuscript. Special thanks to Anton Smirnov for help in preparing the article.

 \bigskip\medskip{REFERENCES}\medskip {\small\begin{enumerate}

 \item
I. Ablimit, G. Zhao, Astrophys. J. {\bf 846}, 10 (2017).

 \item
I. Ablimit, G. Zhao, C. Flynn, and S.A. Bird, Astrophys. J. {\bf 895}, L12 (2020).

 \item
C.P. Ahn, et al., Astrophys. J. Suppl. {\bf 203}, 21 (2012).

 \item
A.T. Bajkova, V.V. Bobylev, Astron. Lett. {\bf 42}, 567 (2016).

 \item
A.T. Bajkova, V.V. Bobylev, Open Astronomy {\bf 26}, 72 (2017a).

 \item
A.T. Bajkova, V.V. Bobylev, Astron. Rep. {\bf 61}, 727 (2017b).

 \item
C. Barber, E. Starkenburg, J.F. Navarro, et al., MNRAS {\bf 437}, 959 (2014).

 \item
G. Battaglia, A. Helmi, H. Morrison, et al., MNRAS {\bf 364}, 433 (2005).

 \item
H. Baumgardt, E. Vasiliev, MNRAS {\bf 505}, 5957 (2021).

 \item
R.A. Benjamin, E. Churchwell, B.L. Babler, et al., PASP {\bf 115}, 953 (2003).

 \item
P. Bhattacharjee, S. Chaudhury, and S. Kundu, Astrophys. J. {\bf 785}, 63 (2014).

\item
J. Binney, S. Tremaine, {\it Galactic Dynamics} (Princeton: Princeton Univ. Press) (1987).

\item
S.A. Bird, X.-X. Xue, C. Liu, et al., MNRAS {\bf 516}, 731 (2022).

\item
J. Bland-Hawthorn, O. Gerhard, Ann. Review Astron. Astrophys. {\bf 54}, 529 (2016).

\item
W.R. Brown, M.J. Geller, S.J. Kenyon, and A. Diaferio, Astron. J. {\bf 139}, 59 (2010).

 \item
V.V. Bobylev, A.T. Bajkova and A.O. Gromov, Astron. Lett. {\bf 43}, 241 (2017).

 \item
J. Bovy, C.A. Prieto, T.C. Beers, et al., Astrophys. J. {\bf 759}, 131 (2012).

 \item
M. Boylan-Kolchin, J.S. Bullock, S.T. Sohn, et al., Astrophys. J. {\bf 768}, 140 (2013).

 \item
S. Buder, et al., MNRAS {\bf 506}, 150 (2021).

 \item
M.T. Busha, P.J. Marshall, R.H. Wechsler, et al., Astrophys. J.  {\bf 743}, 40 (2011).

 \item
T.M. Callingham, M. Cautun, A.J. Deason, et al., MNRAS {\bf 484}, 5453 (2019).

 \item
E. Carlesi, Y. Hoffman, and N.I. Libeskind, MNRAS {\bf 513}, 2385 (2022).

 \item
M. Cautun, C.S. Frenk, R. van de Weygaert, MNRAS {\bf 445}, 2049 (2014).

 \item
M. Cautun, A. Benitez-Llambay, A.J. Deason, et al., MNRAS {\bf 494}, 4291 (2020).

\item
A.J. Deason, V. Belokurov, N.W. Evans, MNRAS {\bf 425}, 2840 (2012).

\item
A.J. Deason, A. Fattahi, V. Belokurov, et al., MNRAS {\bf 485}, 3514 (2019).

\item
A.J. Deason, D. Erkal, V. Belokurov, et al., MNRAS {\bf 501}, 5964 (2021).

\item
L.-C. Deng, H.J. Newberg, C. Liu, et al., Res. Astron. Astrophys. {\bf 12}, 735 (2012).

\item
G.M. Eadie, W.E. Harris, and L.M. Widrow, Astrophys. J. {\bf 806}, 54 (2015).

\item
G.M. Eadie, W.E. Harris, Astrophys. J. {\bf 829}, 108 (2016).

\item
G. Eadie, M. Juri\'c, Astrophys. J. {\bf 875}, 159 (2019).

 \item
A.-C. Eilers, D.W. Hogg, H.-W. Rix, and M.K. Ness, Astrophys. J. {\bf 871}, 120 (2019).

 \item
D.J. Eisenstein, D.H. Weinberg, E. Agol, et al., Astrophys. J. {\bf 142}, 72 (2011).

 \item
G. Fragione, A. Loeb, New Astron. {\bf 55}, 32 (2017).

\item
T.K. Fritz, A. Di Cintio, G. Battaglia, et al., MNRAS {\bf 494}, 5178 (2020).

 \item
Gaia Collab., T. Prusti, et al.,  Astron. Astrophys. {\bf 595}, A1 (2016). 

 \item
Gaia Collab., A.G.A. Brown, et al., Astron. Astrophys. {\bf 616}, 1 (2018). 

 \item
Gaia Collab., A.G.A. Brown, et al., Astron. Astrophys. {\bf 649}, 1 (2021). 

 \item
Gaia Collab., A. Vallenari, et al., arXiv: 2208.0021 (2022).

\item
O.Y. Gnedin, W.R. Brown, M.J. Geller, and S.J. Kenyon, Astrophys. J. Lett. {\bf 720}, L108 (2010).

\item
S.L.J. Gibbons, V. Belokurov, and N. W. Evans, MNRAS {\bf 445}, 3788 (2014).

\item
R.E. Gonz\'alez, A.V. Kravtsov, and N.Y. Gnedin, Astrophys. J. {\bf 793}, 91 (2014).

\item
Y. Huang, X.-W. Liu, H.-B. Yuan, et al., MNRAS {\bf 463}, 2623 (2016).

\item
P.R. Kafle, S. Sharma, G.F. Lewis, and J. Bland-Hawthorn,
 Astrophys. J. {\bf 761}, 98 (2012).

\item
P.R. Kafle, S. Sharma, G.F. Lewis, and J. Bland-Hawthorn,
 Astrophys. J. {\bf 794}, 59 (2014).

\item
I.D. Karachentsev, O.G. Kashibadze, D.I. Makarov, and R.B. Tully,
 MNRAS {\bf 393}, 1265 (2009).

\item
E.V. Karukes, M. Benito, F. Iocco, et al.,
 J. Cosmology and Astroparticle Physics {\bf 05}, 33 (2020).

\item
A.H.W. K\"upper, E. Balbinot, A. Bonaca, Astrophys. J. {\bf 803}, 80 (2015).

\item
Y.-S. Li, S.D.M. White, MNRAS {\bf 384}, 1459 (2008).

\item
Z.-Z. Li, Y.-Z. Qian, J. Han, et al., Astrophys. J. {\bf 894}, 10 (2020).

\item
H. Li, F. Hammer, C. Babusiaux, et al., Astrophys. J. {\bf 916}, 8 (2021).

\item
M. Libralato, A. Bellini, L.R. Bedin, et al., Astrophys. J. {\bf 854}, 45 (2018).

\item
S.R. Majewski, R.P Schiavon, P.M. Frinchaboy, et al., Astron. J. {\bf 154}, 94  (2017).

\item
K. Malhan, R.A. Ibata, MNRAS {\bf 486}, 2995 (2019).

\item
P.J. McMillan, MNRAS {\bf 414}, 2446 (2011).

\item
P.J. McMillan, MNRAS {\bf 465}, 76 (2017).

\item
M. Miyamoto, R. Nagai, Publ. Astron. Soc. Japan {\bf 27}, 533 (1975).

\item
G. Monari, B. Famaey, I. Carrillo, et al., Astron. Astrophys. {\bf 616}, L9 (2018).

\item
J.F. Navarro, C.S. Frenk, and S.D.M. White, Astrophys. J. {\bf 490}, 493 (1997).

\item
L. Necib, T. Lin, Astrophys. J. {\bf 926}, 189 (2022).

\item
J. Pe\~narrubia, Y.-Z. Ma, M.G. Walker, and A. McConnachie, MNRAS {\bf 443}, 2204 (2014).

\item
T. Piffl, C. Scannapieco, J. Binney, Astron. Astrophys. {\bf 562}, 91 (2014).

\item
H.C. Plummer, MNRAS {\bf 71}, 460 (1911).

\item
L. Posti, A. Helmi, Astron. Astrophys. {\bf 621}, 56 (2019).

\item
V. Rashkov, A. Pillepich, A.J. Deason, et al., Astrophys. J. Lett. {\bf 773}, L32 (2013).


 \item
M.F. Skrutskie, R.M. Cutri, R. Stiening, et al., Astrophys. J. {\bf 131}, 1163 (2006).

 \item
M.C. Smith, G.R. Ruchti, A. Helmi, MNRAS {\bf 379}, 755 (2007).

 \item
Y. Sofue, Publ. Astron. Soc. Japan {\bf 61}, 153 (2009).

 \item
Y. Sofue, Publ. Astron. Soc. Japan {\bf 64}, 75 (2012).

\item
S.T. Sohn, G. Besla, R.P. van der Marel, et al., Astrophys. J. {\bf 768}, 139 (2013).

\item
S.T. Sohn, L.L. Watkins, M.A. Fardal, et al., Astrophys. J. {\bf 862}, 52 (2018).

\item
G. Sun, Y. Wang, C. Liu, et al., Res. Astron. Astrophys. {\bf 23}, 5013 (2023).

\item
M. Steinmetz, et al., Astron. J. {\bf 132}, 1645 (2006). 

\item
E. Vasiliev, MNRAS {\bf 484}, 2832 (2019).

\item
R.P. van der Marel, M. Fardal, G. Besla, et al., Astrophys. J. {\bf 753}, 8 (2012).

 \item
S. Wang, X. Chen, R. de Grijs, et al., Astrophys. J. {\bf 852}, 78 (2018).

 \item
W. Wang, J. Han, M. Cautun, et al., Science China Physics, Mechanics,
  and Astronomy {\bf 63}, 109801 (2020).

\item
J. Wang, F. Hammer, and Y. Yang, MNRAS {\bf 510}, 2242 (2022).

\item
L.L. Watkins, N.W. Evans, and J.H. An, MNRAS {\bf 406}, 264 (2010).

\item
L.L. Watkins, R.P. van der Marel, S.T. Sohn, and N.W. Evans, Astrophys. J. {\bf 873}, 118 (2019).

\item
A.A. Williams, N.W. Evans, MNRAS {\bf 454}, 698 (2015).

\item
A.A. Williams, V. Belokurov, A.R. Casey, and N.W. Evans, MNRAS {\bf 468}, 2359 (2017).

 \item
E. Wright, et al., Astron. J. {\bf 140}, 1868 (2010).

\item
D.G. York, et al., Astron. J. {\bf 120}, 1579 (2000).

\item
X.X. Xue, H.W. Rix, G. Zhao, et al., Astrophys. J. {\bf 684}, 1143 (2008).

\item
D. Zaritsky, H. Courtois, MNRAS {\bf 465}, 3724 (2017).

\item
M. Zhai, X.-X. Xue, L. Zhang, et al., Res. Astron. Astrophys.  {\bf 18}, 113 (2018).

\item
Y. Zhou, X. Li, Y. Huang, and H. Zhang, arXiv: 2212.10393 (2022).

 \end{enumerate}
 }
\end{document}